\documentstyle[epsf,12pt]{article}
\begin{document}
\begin{titlepage}
\begin{center}
\vspace*{3cm}

{\Large
\bf Entropy in hadron-hadron collisions          
 \\ 
}

\vspace{2cm}
{\large
K. Fia{\l}kowski\footnote{e-mail address: uffialko@thrisc.if.uj.edu.pl},
R. Wit\footnote{e-mail address: wit@thrisc.if.uj.edu.pl}
}
\vspace{1cm}

{\sl M. Smoluchowski Institute of Physics\\ Jagellonian University \\

30-059 Krak{\'o}w, ul.Reymonta 4, Poland}

\vspace{3cm}

\begin{abstract}
We analyse the entropy properties in the proton - proton 1800 GeV 
events from the PYTHIA/JETSET Monte Carlo generator following a recent 
proposal 
concerning the measurement of entropy in multiparticle systems.   
The dependence on the number of bins and on the size of the phase-space 
region is investigated. Our results may serve as a reference sample for 
experimental data from hadron-hadron and heavy ion collisions.

\end{abstract}

\end{center}



\vspace{1cm}

\noindent

 13 July, 2000 \\

\end{titlepage}

\section{Introduction}
\par
In a recent series of papers Ref. \cite{BCW}, \cite{BC1}, \cite{BC2}
a specific proposal was presented for entropy measurement of multiparticle systems 
created in high-energy collisions. The proposal should be important for
the analysis of the forthcoming RHIC experiments where it may help in the separation
of a possible signal from quark-gluon plasma (QGP).
However, this proposal  is not merely restricted to systems 
with a very large number of particles. Applying it to other multiparticle systems, e.g.
originating from hadron-hadron collisions, one may get useful reference data for the 
discussion of the thermodynamic equilibrium and other properties of such systems.
\par
In this note we use the PYTHIA/JETSET event generator \cite{SJO} to create samples 
of multiparticle states and analyse them according to the proposal mentioned above.
In the next section we remind shortly  the procedure presented in 
Ref. \cite{BC2} and specify the process and variables used in the analysis.
The results are presented in the third section. The last section contains  
a discussion of the results including some conclusions and perspectives.

\section{Procedure and variables}
We generate samples of $10^5$ or $10^6$ events of $pp$ collisions at $1800 ~GeV$    
CM energy, the highest energy available yet for hadron-hadron collisions. 
This ensures a relatively high particle density leaving the possibility for 
comparison with experimental data. For each  event the phase space region 
of a few units in rapidity (in the central region) and $p_T^2$ restricted 
to less than $0.4 ~GeV^2/c^2$ is used. 
\par
To calculate the entropy we "discretize" each event. For definiteness 
we are using bins in $p_T^2$; binning in rapidity leads to similar results.
The $p_T^2$ range is divided 
into $M$ bins, and the number of particles in each bin $m_i, ~i=1,...,M$ is 
recorded. Now it is possible to calculate the Shannon entropy from the 
standard definition 

\begin{equation}
\label{eq1}
S = - \sum _j p_j \log p_j
\end{equation}
where $p_j$ denotes the probability to obtain any specific configuration of numbers 
$\{m_i\}$. Obviously, $p_j = n_j/N$ where $n_j$ is the number of events providing 
such configuration and $N$ is the global number of events.
\par 
 However, in the proposal \cite{BC2} the calculation of entropy is  
 performed in a different way for the reasons to be discussed later. First, 
 one calculates total numbers of observed 
 coincidences of $k$ configurations $N_k$
\begin{equation}
\label{eq2}
N_k = \sum _j n_j(n_j -1)...(n_j - k +1).
\end{equation}
Then the coincidence probability of $k$ configurations is given by 

\begin{equation}
\label{eq3}
C_k  = \frac{N_k}{N(N-1)...(N-k+1)}.
\end{equation}  
These probabilities are used to calculate Renyi entropies as  
\begin{equation}
\label{eq4}
H_k  = -\frac{\log C_k}{k-1}
\end{equation}
instead of calculating them directly from the following 
definition 
\begin{equation}
\label{eq5}
H_k  = \sum _j (p_j)^k.
\end{equation} 
We will comment later on the Renyi entropy values obtained by these two methods.
\par
The Shannon entropy is formally equal to the limit of Renyi entropies $H_k$ as 
$k \rightarrow 1$ and can be obtained by extrapolation. Obviously $N_1 = N$, $C_1 = 1$,
and this extrapolation cannot be done just by putting $k=1$ in formula (\ref{eq4}).
It was suggested \cite{BC1} to use for the extrapolation a formula 
\begin{equation}
\label{eq6}
H_k  = a\frac{\log k}{k-1}+a_0 + a_1(k-1)+ a_2(k-1)^2 + ...,
\end{equation}
where the number of terms is determined by the number of measured Renyi 
entropies. Usually it is enough to use $H_k$ for $k =2,3,4$. Other 
extrapolations can also be used, e.g. 
\begin{equation}
\label{eq61}
H_k  = a_0 + a_1k^{-1}+ a_2k^{-2} + ...,
\end{equation}
and should be compared with the one 
presented here to estimate the extrapolation accuracy. We comment on this later on.
\par
It was suggested  that for a system close to equilibrium and small bins 
the entropy should grow logarithmically with the number of bins
\begin{equation}
\label{eq7}
H_k(lM)  = H_k(M) + \log l ~~\Rightarrow ~~ S(lM) = S(M) + \log l. 
\end{equation}
Another expected feature is additivity: for entropies measured in a phase-space 
region $R$,
which is the sum of two regions $R_1$ and $R_2$, we should observe 
\begin{equation}
\label{eq8}
H_k(R)  = H_k(R_1) + H_k(R_2) ~~\Rightarrow ~~S(R) = S(R_1) + S(R_2). 
\end{equation}
We  check these features by choosing different numbers of bins in $p_T^2$ 
and different ranges of rapidity.
\par 
As we have seen, for all our calculations we need $n_j$, the numbers of events
 providing specific configurations of numbers of particles in bins ${m_i}$.
It may be difficult to record all $n_j$ since the number of different 
possible configurations grows quickly with the number of bins and particles. 
If for each of M bins the number of particles may change within an interval length
 $L$ the number of a priori possible 
configurations is $L^M$. This is a rather 
big   number even for  moderate values of $L$ and $M$.
\par
However, we need to know only the values of $n_j$ and not the 
form  of the configurations corresponding to each value of 
$n_j$. Therefore it is really not necessary to make big computer 
memory reservations. Instead of initializing a big matrix with all elements 
equal to zero and filling it gradually with generated events, we define 
$n_j$ only 
for those configurations which actually appear in generated events.
\par
Still, the registration of all $n_j$-s and consecutive calculations may be 
quite time consuming. In our case we have found that the computing time 
may become prohibitive for $10^6$ events and  9 $p_T^2$ bins. Therefore it
 is important to find the lowest possible number of events for which the 
 results become stable.

\section{Results}
\par 
To start discussing any results one should know how reliable they are and 
what is their uncertainty. Therefore we estimated first the 
dependence of entropy values on the generated number of events. We checked 
that the results for $10^5$ and $10^6$ event samples differ most for largest
number of bins and largest rapidity range $\mit \Delta y$. We show 
this effect in Tab.1, 
where  the values of Renyi and Shannon entropies are presented for 
9 bins in $p_T^2$ and different values of $\mit \Delta y$ for these two 
samples of events. Shannon entropy is 
calculated by extrapolation and from the direct definition (1). 
For smaller number of bins all differences are smaller, but
the pattern is the same.\\

TAB.1. 
Entropy values for 9 bins in $p_T^2$ for changing 
 $\mit \Delta y$ from $10^6~(10^5)$ events. 
{\begin{center}
{ 
\begin{tabular}{||c|c|c|c|c|c|c}
\hline
\hline
\multicolumn {1}{||c|}{}&
\multicolumn {5}{|c|}{$\Delta y$} \\
\hline
\hline
 Entropy &  1 & 2 &  3&  4 & 6 \\
\hline
\hline
$H_4$ & 3.67 (3.66) &5.84 (5.84) &7.47 (7.51) &8.85 (8.90) &11.03 (11.07) \\
\hline
$H_3$& 3.99 (3.98) &6.27 (6.28) &
8.00 (8.03) &9.43 (9.46)  & 11.63 (11.66)   \\
\hline
$H_2$ & 4.68 (4.68) & 7.19 (7.20) &9.03 (9.04) &10.51 (10.52) &12.69 (12.70) \\
  \hline
$S^{ex}$ &6.44 (6.45) &9.47 (9.47) &11.53 (11.51)& 13.03 (13.04) &15.07 (15.07) \\
\hline
$S^{df}$ & 6.76 (6.64) &9.46 (9.04) &11.06 (10.23) &12.07 (10.88)&13.12 (11.36) \\

\hline
\hline
\end{tabular}
}
\end{center}
}

The same results are shown in Fig.1. However, for the sake of transparency  only
the values of Shannon entropies and second Renyi entropies are presented. 

\begin{figure}[h]
\centerline
{
\epsfxsize=7cm
\epsfbox{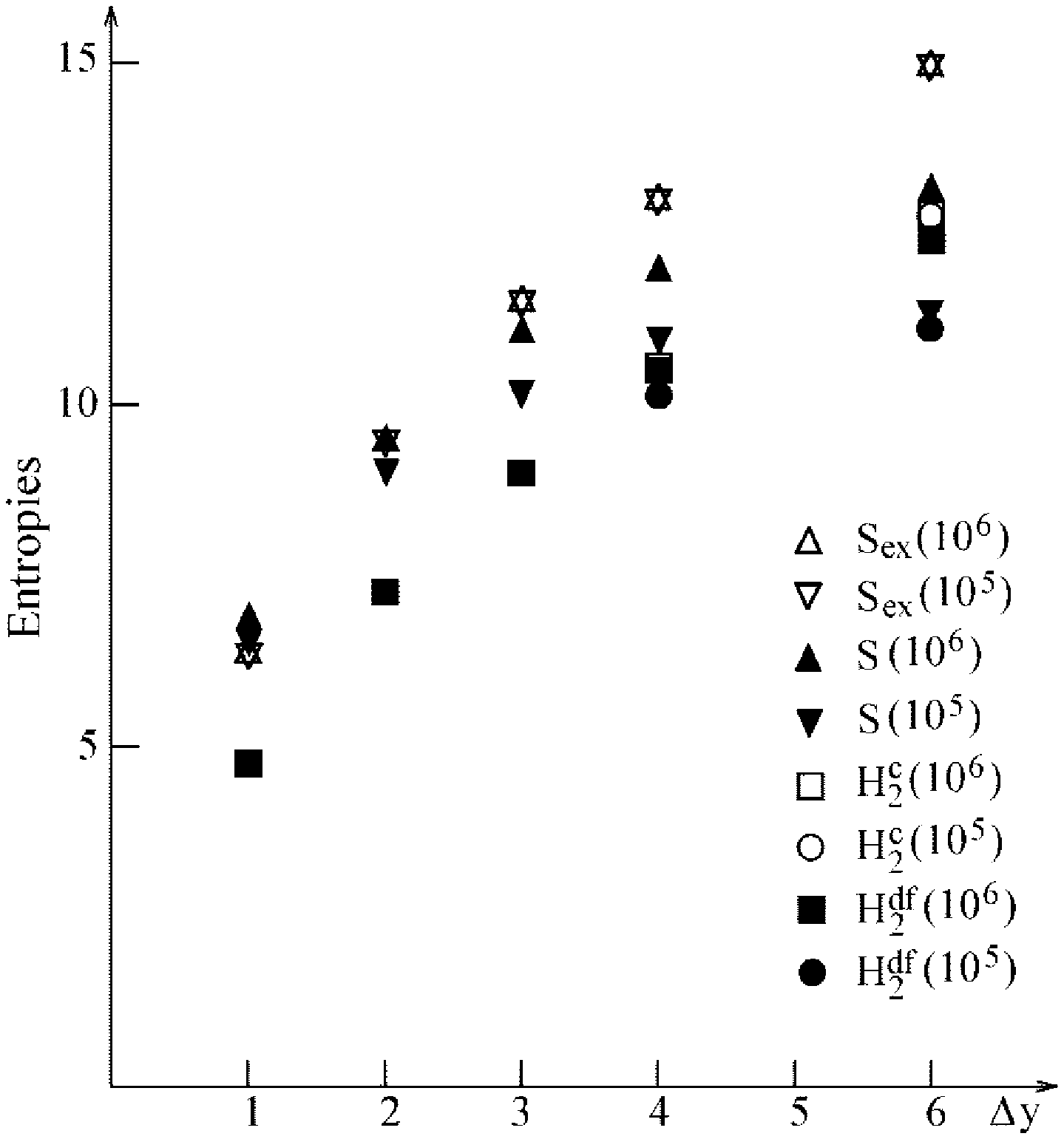}
}
\caption{\label{fig1} {\sl \small Shannon entropy calculated for 9 bins in $p_T^2$
for $10^5$ events (inverted 
triangles) and for $10^6$ events (triangles) from the definition (\ref{eq1}) 
(black 
symbols) and by extrapolation of formula (\ref{eq6}) (open symbols) as a 
function of rapidity range $\mit \Delta y$. 
Second Renyi 
entropy is also shown for $10^5$ events (circles) and $10^6$ events 
(squares) calculated from 
definition (\ref{eq5}) (black symbols) and from the coincidence 
probabilities (\ref{eq4}) (open symbols). }}

\vspace{0.4cm}
\end{figure}

\par
We see that for narrow ranges of rapidity $\mit \Delta y$ the values of entropy
 for $10^5$ and $10^6$ events are very similar and the differences grow with 
 $\mit \Delta y$. The most striking effect is that these differences stay always 
 small when Shannon entropy $S$ is calculated by the extrapolation from Renyi
 entropies $H_k$ to $k=1$ according to formula (6), whereas the values calculated
 directly from the definition (1) differ really strongly for two samples 
 at widest rapidity ranges. In fact, the entropy values calculated
 for $10^5$ events from the definition (1) 
 seem to saturate at the level of $11.5$, which is close to $log 10^5$.
\par
This confirms that the method proposed in \cite{BCW}, \cite{BC1} is indeed much better 
than the direct measurment of Shannon entropy (unless the number of particles in the bin 
is too small).
For this method it is possible to calculate the Shannon 
entropy reliably even  for modest samples of events. In the following we show only 
the values obtained by the extrapolation procedure. We checked that the results for 
the second extrapolation (7) are the same within $2\%$ accuracy.
\par
For the second (and further) Renyi entropies the results for two samples never 
differ too much. The difference is still smaller if they are calculated by
the advised procedure from coincidence probabilities (4) and not directly from 
the definition (5). Further results shown use always this procedure.
\par
Before testing the additivity of entropy (relation(8))
we perform a simple exercise. Since it was suggested that additivity 
may be broken by correlation effects, we checked if the short-range correlations
are relevant. To this purpose we calculated the entropies for the same number of
bins in $p_T^2$ using the rapidity range ${\mit \Delta} y = 2$ centered at $CM$ rapidity
zero in "one piece" ($- 1< y <1$) and in two intervals of width 1 separated by a gap of
two units ($-2 < y <-1$ and $1 < y <2$). As seen in Fig.2, the results are barely 
distinguishable, which
shows that the short range correlation effects are negligible for our discussion.
 
\begin{figure}[h]
\centerline{
\epsfxsize=7cm
\epsfbox{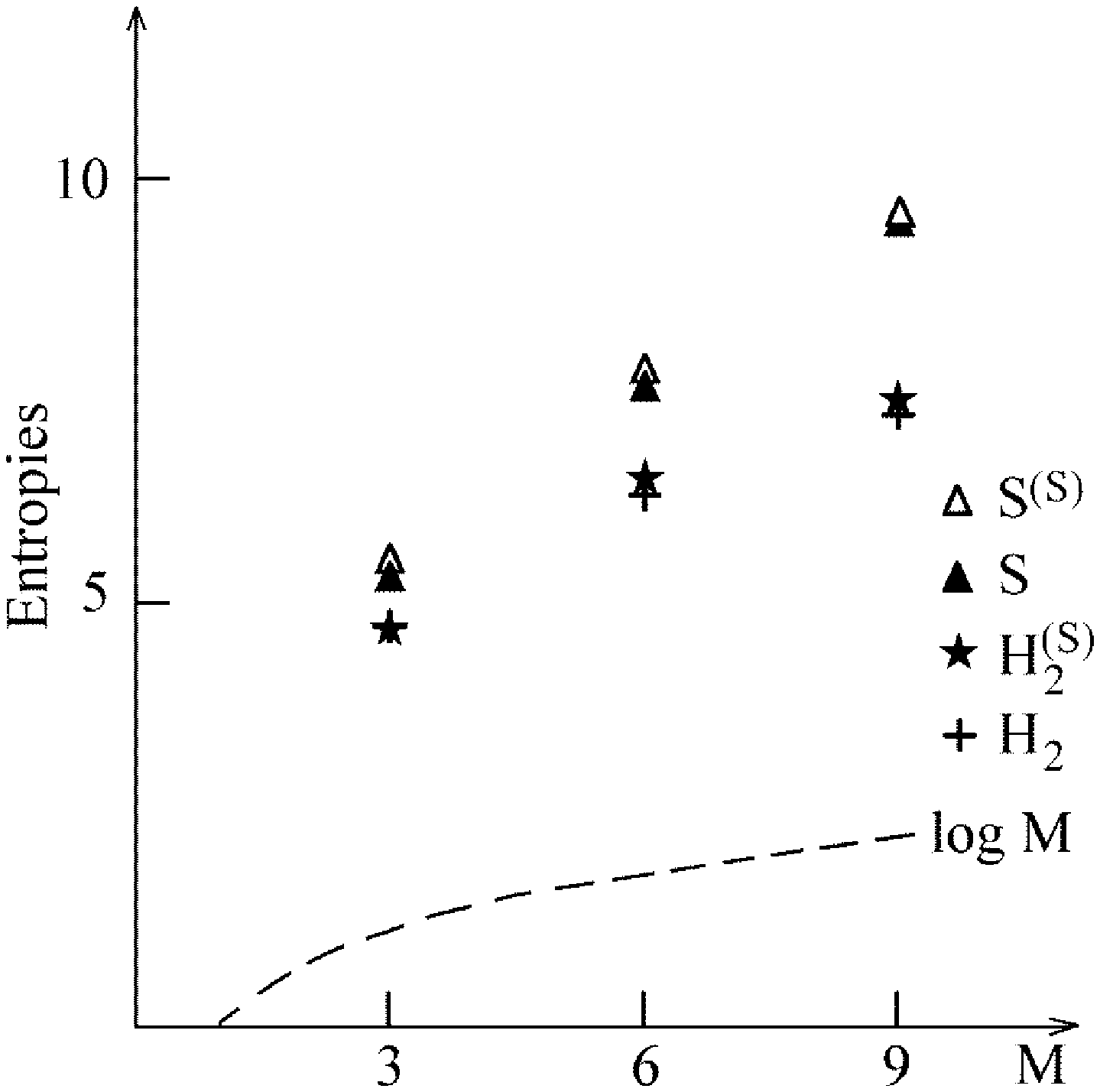}
}
\caption{\label{fig2} {\sl \small Shannon entropy for compact (black 
triangles)
and separated (open triangles) phase space regions in rapidity of two 
units width as a function
of number of bins $M$. Second Renyi entropy is also shown (crosses and stars,
 respectively). }}
\vspace{0.4cm}
\end{figure}

\par
The dependence of the entropy on the number of bins seen in this figure
seems to be significantly stronger than that predicted by eq. (7): the
$\log M$ curve is shown on the bottom of the Fig.2 for comparison.
\par
The irrelevance of short-range correlations allows us to test additivity 
simply by plotting the dependence of 
entropy values on the width of rapidity range $\mit \Delta y$. Entropy  
should be proportional to $\mit \Delta y$ (at least in the central region, 
where the rapidity 
distribution is approximately flat; our values of $\mit \Delta y$ correspond
always to this region). We check it for two choices of bins: of equal width
in $p_T^2$ (as in all other figures) and for the same number of bins, 
the same range of $p_T^2$, but bin sizes defined by a requirement of 
approximately equal average multiplicities. The results are shown in Fig.3 
for 6 bins; for other numbers of bins the pattern is the same. 

\begin{figure}[h]
\centerline{
\epsfxsize=7cm
\epsfbox{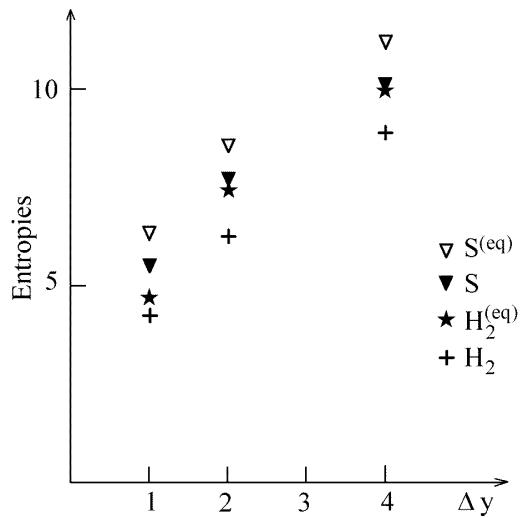}
}
\caption{\label{fig3} {\sl \small Shannon entropy for 6 bins in $p_T^2$ 
of equal 
width (black triangles) and of equal multiplicity (open triangles) as a 
function of rapidity range $\Delta y$. Second Renyi entropy is also shown  
(crosses and stars, respectively).   }}
\vspace{0.4cm}
\end{figure}

\par
We see that the results for two binning procedures differ just by shifting 
the entropy values; the dependence on rapidity range is the same and in both
cases it is definitely weaker than linear. Thus there is no additivity 
in the sense
of eq. (8), which turn suggests that there is no thermal equilibrium 
in the
process under investigation.
\par
Finally, we test the dependence of entropies on the number of bins when the
average multiplicities per bin remain unchanged (i.e. we increase the rapidity
range proportionally to the number of bins in $p_T^2$ keeping thus the "bin
volume" $\Delta V = \Delta p_T^2 \cdot \Delta y$ constant). The results are 
shown in Fig.4. As one sees, the dependence is in this case approximately 
linear.

\begin{figure}[h]
\centerline{
\epsfxsize=7cm
\epsfbox{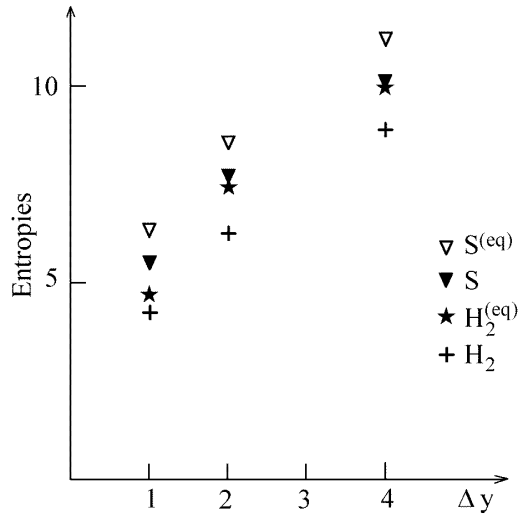}
}
{\sl
\caption{\label{fig4} {\small Shannon entropy for equal width bins in 
$p_T^2$
 of constant bin volume $\Delta V = \Delta y \cdot \Delta p_T^2$ as a 
 function of number of bins $M$. Black 
 (open) triangles are for $\Delta V = 0.133 ~GeV^2$ ($0.267 ~GeV^2$). Second 
 Renyi entropy is also shown (crosses and stars, respectively). }}
 }
\vspace{0.4cm}
\end{figure}

\section{Conclusions and outlook}
\par 
We have calculated Shannon entropies for the final states from the $pp$ collisions
at $1800 ~GeV$ CM energy using the PYTHIA/JETSET event generator. We found 
that for the procedure extrapolating Renyi entropies it is enough to generate 
$10^5$ events to get numerically stable results. 
\par
We have tested the conjecture that entropy is additive, i.e. that entropy measured 
in a phase-space region $R$ which is the sum of two regions $R_1$ and $R_2$ 
is just a sum of entropies  measured in these two regions. Our results do not 
confirm this conjecture; the increase of entropy with the size of the phase-space 
region is slower than linear.  This is may be regarded as the effect of 
correlations. We show that it is dominated by long range correlations; 
the results for two adjacent regions and separated regions
are almost the same.
\par
We have also investigated the dependence of entropy on the number of bins. It seems to
 be stronger than the expected logarithmic, perhaps due to a small number of bins.
If we keep the average multiplicity per bin unchanged and increase both the number of 
bins and the size of the relevant phase-space region, we find an approximately linear 
increase of entropy.
\par
Our investigation shows that it is feasible to perform a program proposed 
in Refs. \cite{BCW}, \cite{BC1} and \cite{BC2} for experimental data. The procedure 
will be the same as for our samples of generated events. We have shown that for 
 hadron-hadron collisions the results are stable already for $10^5$ events. Obviously,
 for the high multiplicity heavy ion collisions one should take much smaller bins 
 to have comparable multiplicities; otherwise one should check again  the stability 
conditions.
\par
The results presented above may also serve as a reference sample for the 
experimental data resulting both from hadron-hadron and heavy ion collisions. 
Since the used generator does not assume any thermodynamical equilibrium, 
the observed similarities and differences may help in the discussion concerning 
the presence of equilibrium in data.
\par
It would be useful to perform a similar analysis for different choices of variables 
and for different generators,  in particular for those which are dedicated for heavy 
ion collisions. 

\section{Acknowledgements}
We would like to thank A. Bia{\l}as and W. Czy\.z for reading the manuscript.
The financial support of KBN grants \# 2 P03B 086 14 and \# 2P03B 010 15 
is gratefully acknowledged.  One 
of us (RW) is grateful for a partial financial support 
by the KBN grant \# 2 P03B 019 17 and the other one (KF) for the support of FNP 
(subsydium FNP 1/99 granted to A. Bia{\l}as).

\end{document}